\documentclass[sigconf,screen]{acmart}

\copyrightyear{2024}
\acmYear{2024}
\setcopyright{rightsretained}
\acmConference[CHI PLAY Companion '24]{Companion Proceedings of the Annual Symposium on Computer-Human Interaction in Play}{October 14--17, 2024}{Tampere, Finland}
\acmBooktitle{Companion Proceedings of the Annual Symposium on Computer-Human Interaction in Play (CHI PLAY Companion '24), October 14--17, 2024, Tampere, Finland}\acmDOI{10.1145/3665463.3678820}
\acmISBN{979-8-4007-0692-9/24/10}

\makeatletter
\gdef\@copyrightpermission{
  \begin{minipage}{0.2\columnwidth}
   \href{https://creativecommons.org/licenses/by-sa/4.0/}{\includegraphics[width=0.90\textwidth]{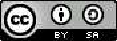}}
  \end{minipage}\hfill
  \begin{minipage}{0.8\columnwidth}
   \href{https://creativecommons.org/licenses/by-sa/4.0/}{This work is licensed under a Creative Commons Attribution-ShareAlike International 4.0 License.}
  \end{minipage}
  \vspace{5pt}
}
\makeatother

\usepackage{float}
 \usepackage{balance}

\usepackage[shortlabels]{enumitem}

\usepackage{xspace} 
\newcommand{\guide}{\textit{Assessment Guide}\xspace}
\newcommand{\wip}{WIP\xspace} 

\newcommand{\Moss}{\textit{Moss2}\xspace}
\newcommand{\BS}{\textit{Beat Saber}\xspace}
\newcommand{\OW}{\textit{OW2}\xspace}

\usepackage{microtype}  
\newcommand{\codestyle}[1]{\textit{#1}}
\newcommand{\mechanics}[1]{\texttt{\textls[-40]{#1}}}

\usepackage{multirow}
\usepackage{fontawesome}

\newcommand{\deficon}{\faChevronCircleRight\xspace} 
\newcommand{\coricon}{\faChevronRight\xspace} 

\usepackage{colortbl}
\makeatletter
\def\adl@drawiv#1#2#3{%
	\hskip.5\tabcolsep
	\xleaders#3{#2.5\@tempdimb #1{1}#2.5\@tempdimb}%
	#2\z@ plus1fil minus1fil\relax
	\hskip.5\tabcolsep}
\newcommand{\cdashlinelr}[1]{%
	\noalign{\vskip\aboverulesep
		\global\let\@dashdrawstore\adl@draw
		\global\let\adl@draw\adl@drawiv}
	\cdashline{#1}
	\noalign{\global\let\adl@draw\@dashdrawstore
		\vskip\belowrulesep}}
\makeatother

\usepackage{soul}

\usepackage{xcolor,colortbl}
\definecolor{All-gray}{HTML}{dddcdc}
\definecolor{palered}{HTML}{ffadad}
\definecolor{paleorange}{HTML}{FFCA85}  
\definecolor{paleyellow}{HTML}{fdffb6}
\definecolor{palegreen}{HTML}{caffbf}
\definecolor{palecyan}{HTML}{9bf6ff}
\definecolor{paleblue}{HTML}{a0c4ff}   
\definecolor{palepurple}{HTML}{bdb2ff}
\definecolor{palepink}{HTML}{ffc6ff}
\definecolor{palewhite}{HTML}{fffffc}

\usepackage{rotating}
\begin{document}

\title[Deceptive Design in VR Games]{Computer-based Deceptive Game Design in Commercial Virtual Reality Games: A Preliminary Investigation}

\author{Hilda Hadan}
\email{hhadan@uwaterloo.ca}
\orcid{https://orcid.org/0000-0002-5911-1405}
\affiliation{
    \institution{Stratford School of Interaction Design and Business, University of Waterloo}
    \city{Waterloo}
    \country{Canada}
}

\author{Leah Zhang-Kennedy}
\email{lzhangke@uwaterloo.ca}
\orcid{https://orcid.org/0000-0002-0756-0022}
\affiliation{
    \institution{Stratford School of Interaction Design and Business, University of Waterloo}
    \city{Waterloo}
    \country{Canada}
}

\author{Lennart E. Nacke}
\email{lennart.nacke@acm.org}
\orcid{https://orcid.org/0000-0003-4290-8829}
\affiliation{
    \institution{Stratford School of Interaction Design and Business, University of Waterloo}
    \city{Waterloo}
    \country{Canada}
}

\renewcommand{\shortauthors}{Hilda Hadan, Leah Zhang-Kennedy, \& Lennart E. Nacke}

\begin{abstract}

As Virtual Reality (VR) games become more popular, it is crucial to understand how deceptive game design patterns manifest and impact player experiences in this emerging medium.
Our study sheds light on the presence and effects of manipulative design techniques in commercial VR games compared to a traditional computer game. We conducted an autoethnography study and developed a VR Deceptive Game Design Assessment Guide based on a critical literature review. Using our guide, we compared how deceptive patterns in a popular computer game are different from two commercial VR titles. While VR's technological constraints, such as battery life and limited temporal manipulation, VR's unique sensory immersion amplified the impact of emotional and sensory deception. Current VR games showed similar but evolved forms of deceptive design compared to the computer game. We forecast more sophisticated player manipulation as VR technology advances. Our findings contribute to a better understanding of how deceptive game design persists and escalates in VR. We highlight the urgent need to develop ethical design guidelines for the rapidly advancing VR games industry.
\end{abstract}




\maketitle

\section{Introduction}
\label{sec:introduction}
Deceptive game design has been a growing area of concern for player researchers. Literature has identified various deceptive game design patterns used in mobile and computer-based games to manipulate player behaviour for the benefit of game developers and publishers (e.g.,~\cite{fitton2019F2P,roffarello2023defining,geronimo2020UI,karlsen2019exploited}). These practices designed to manipulate player engagement and monetization through psychological and emotional tactics can escalate frustration and confusion~\cite{zagal2013dark,gray2021enduser,maier2019dark}, and damage game publisher reputations~\cite{hhadan2024ow2}.
Deceptive design research has primarily focused on games on traditional 2D interfaces (e.g.,~\cite{zagal2013dark,fitton2019F2P,karlsen2019exploited}). However, commercially available Virtual Reality (VR) technology opens a new frontier for immersive manipulative tactics.
Spatial and realistic displays in VR could amplify the impact and subtlety of deceptive designs~\cite{hadan2024deceived,krauss2024what}, impairing users' ability to critically evaluate content~\cite{cummings2022all}. As commercial game development shifts to VR platforms, understanding how deceptive practices adapt to immersive technology is paramount. 

Although recent studies have explored how deceptive design could work in immersive environments, these explorations have mainly focused on hypothetical scenarios from literature, experts predictions, and controlled laboratory experiments (e.g.,~\cite{hadan2024deceived,krauss2024what,su2022perception}). These studies focused broadly on XR technology and neglected the specific contextual factors of VR games. 
The impact of deceptive design on players within immersive VR environments remains largely unexplored. Our research addresses this gap by investigating deceptive design patterns in commercially available VR games, and comparing them to traditional game implementations. We analyze how these patterns evolve across modalities, with a focus on their potential impact in VR and on player experience. Our overarching Research Questions (RQs) are: 

\begin{enumerate}[label=\textbf{RQ\arabic*:}]
    \item What deceptive design patterns are currently implemented in commercial VR games?
    \item How do deceptive game designs vary in their manifestations in VR games and PC games?
\end{enumerate}

To address these RQs, we are conducting a series of analyses, including a literature synthesis to construct a \textit{VR Deceptive Game Design Assessment Guide} (referred to as the \guide) that comprehensively captures deceptive design patterns in VR games. We are also conducting autoethnography research to identify these patterns in various VR games and to compare new manifestations in VR to those in recent PC games.
In this work-in-progress (\wip), we report the first stage of our research, which includes an initial version of the \guide synthesized from the literature, a preliminary seven-week autoethnography study, and a comparative analysis of deceptive design patterns identified in two commercial VR games (see~\autoref{fig:screenshots}, left and center): \textit{Moss: Book II} (\Moss), a VR puzzle adventure game; 
and \BS, a VR rhythm game. These games are selected because of their critical acclaim in the VR player community and significant player bases\footnote{As of May 10, 2024, \BS has about 862,000 active players; \textit{Moss Book: II} has been played by 33,000 players. See~\url{https://playtracker.net/insight/game/5}.}. We chose a third cross-platform computer game, \textit{Overwatch2} (\OW), a first-person team-based shooter (see~\autoref{fig:screenshots}, right) as an exemplar PC game to compare to VR games because our previous research~\cite{hhadan2024ow2} had identified various deceptive game design mechanisms in \OW
and critical player reviews\footnote{As of May 4, 2024, \OW holds the sixth position on the list of worst games on Steam with 82\% of negative reviews, resulted in an overall rating of ``overwhelmingly negative''. Currently, \OW  See:~\url{https://store.steampowered.com/app/2357570/Overwatch_2/} and~\url{https://steam250.com/bottom100}}. While deceptive design varies across game types, genres, and mechanics, we have focused on examining and contrasting these patterns in a select sample of exemplary games in our preliminary investigation presented in this \wip. This approach allows us to explore how these patterns uniquely manifest and impact VR players more than traditional computer games.

The lead researcher---experienced in deceptive design research and VR gaming applications---extensively played \OW (20 hours in 6 days), followed by \Moss (8 hours in 9 days) then \BS (18 hours in 9 days)  to systematically document detailed diary entries that reflect on moments where game mechanics appeared to use deceptive patterns and produce manipulative experiences. The length of play between the three games is determined by the principle of saturation when no new deceptive design patterns emerged during gameplay. The autoethnographic approach allowed us to actively engage as players and experience deceptive design first-hand, capture player frustrations and excitements, and systematically document these experiences over time~\cite{rapp2018autoethnography}. 
Next, we developed an initial version of the \textit{VR Deceptive Game Design Assessment Guide} across eight themes that consist of 71 distinct deceptive designs from the literature to deductively analyze diary entries (see~\autoref{tab:initial-codebook}). This \guide was constructed by synthesizing existing classifications (e.g.,~\cite{gray2024ontology,zagal2013dark,king20233d,hadan2024deceived}) that comprehensively captured problematic design practices from the academic and regulatory space~\cite{gray2024ontology}, as well as from games~\cite{zagal2013dark}, immersive environments~\cite{hadan2024deceived}, and player perspectives~\cite{king20233d}.

Our preliminary research outcomes make three \textbf{contributions}: \textit{First}, we present a \textit{VR Deceptive Game Design Assessment Guide} that builds on established classifications to serve as a foundation for analyzing deceptive design in VR games, and informing future researchers, VR designers, and policymakers by providing a standardized tool for identifying and categorizing deceptive tactics in VR games. 
\textit{Second}, our research uses an autoethnography method to provide valuable insights into players' lived experiences from a first-person perspective. \textit{Third}, our preliminary investigation revealed various manifestations of deceptive designs across modalities. Our findings suggest that, although the limitations of VR device battery life clashed with tactics driving extensive playtime, VR's unique features amplified the impact of emotional and sensory manipulations and presented a potential for more sophisticated manipulation in future VR games.

\section{Related Work and How It Informed Our Research}
\label{sec:related-work}


\subsection{Deceptive Design in Games and Immersive Environment and The Role of Player Perception}

Deceptive design that impairs users' ability to make informed decisions~\cite{mathur2021makes,gray2023towards,EUDSA2022,Brignull2023book}, is widely used on websites~\cite{mathur2019dark,gray2018dark,gunawan2021comparative}, mobile apps~\cite{lewis2014irresistible,fitton2019F2P,gunawan2021comparative}, and social networks~\cite{mildner2023defending,mildner2023engaging}. 
In games, deceptive designs can benefit game developers but negatively impact player experiences~\cite{zagal2013dark}. The literature has classified deceptive designs for psychological manipulation~\cite{zagal2013dark}, attention capture~\cite{roffarello2023defining}, and extended playtime and money spending~\cite{dillon2020digital,geronimo2020UI,karlsen2019exploited,zagal2013dark}. Studies on 3D interaction~\cite{Greenberg2014proxemic,king20233d} and Extended Reality (XR)\footnote{Extended Reality (XR) describes immersive technologies such as Virtual Reality (VR), Augmented Reality (AR), and Mixed Reality (MR).} literature~\cite{hadan2024deceived} indicate that XR's unique features, such as spatial displays, multi-modality and realistic experiences, real-world sensory blocking, and sensor data, create new opportunities for deceptive design
that amplifies user manipulation~\cite{hadan2024deceived,krauss2024what}. \citet{gray2024ontology} synthesized a comprehensive ontology of 64 deceptive designs from widely adopted regulatory and academic taxonomies.

Deceptive designs exploit internal \textit{human rationality and cognitive limitations} and external \textit{social influence} to manipulate human decision-making~\cite{simon1997models,waldman2020cognitive,thaler2009nudge}. The effectiveness of deceptive design varies based on users' literacy~\cite{zagal2013dark,geronimo2020UI,luguri2021shining}, encounter frequency, perceived trustworthiness, level of frustration, misleading behaviour and UI appearance~\cite{m2020towards}. Player perceptions also play a role, with some feeling ``interrupted'' or questioning game fairness, while others might appreciate the guidance~\cite{zagal2013dark,frommel2022daily,fitton2019F2P,freeman2022pay} or accept it due to service dependency~\cite{maier2019dark}. Given this complexity, literature suggests that user-centred approaches are necessary for identifying design practices that are not straightly unethical~\cite{gray2021enduser,gray2023mapping,hhadan2024ow2}. 

In our preliminary investigation, we built upon~\citet{gray2024ontology}'s comprehensive ontology which was grounded in foundational literature and taxonomies in the field (e.g.,~\cite{Brignull2010deceptive,gray2018dark,mathur2021makes}). We further incorporated insights from~\citet{zagal2013dark} on game-specific and~\citet{hadan2024deceived} on XR-specific deceptive designs, and~\citet{king20233d} on player perception in 3D interfaces. While~\citet{zagal2013dark}'s framework lacked empirical data, it remains a ``fruitful'' starting point for game-focused deceptive design research~\cite[p.~2]{deterding2020against}. Our autoethnography methodology addresses this gap and follows the recommendations by incorporating player experiences and perceptions from a first-person perspective. We present how these studies informed our process of developing our \textit{VR Deceptive Game Design Assessment Guide} in~\autoref{subsec:codebook-development}.

\subsection{Autoethnography as a Methodology in Games Research}

Autoethnography is a common HCI methodology where designers reflect on their extended use of a system to understand its nature and refine its design directions (e.g.,~\cite{rapp2018autoethnography,neustaedter2012autobiographical,boehner2008interfaces}).
In the study of video games, the autoethnography method has been used to study game design elements that enhanced player motivation and engagement~\cite{rapp2017designing,rapp2017games}, shaped gendered experiences~\cite{nardi2010my}, and disrupted players in an AR game~\cite{laato2022balancing}. This method situates researchers as both the ``protagonist'' and observer and enables a first-person engagement and understanding of the game environment~\cite{rapp2018autoethnography}. Other studies have also used this method to study the cultural and social dynamics, such as in-game as well as out-of-the game love affairs~\cite{sunden2012desires}, collaborative play~\cite{nardi2006strangers}, gameplay learning with peers~\cite{nardi2007learning}, players' perception of cheating~\cite{boldi2023legit}, and the emergence of toxic in-game behaviour~\cite{kordyaka2023cycle}. However, to date, and to our knowledge, only one autoethnographic study has been conducted to investigate the impact of deceptive game design on players in computer-based games (i.e.,~\cite{karlsen2019exploited}). 

Literature suggests that XR technology can employ deceptive designs that occur frequently or even constantly over extended periods, making them less noticeable to users~\cite{krauss2024what}, especially when XR devices are used constantly throughout people's daily lives~\cite{hadan2024deceived}. Therefore, we argue that autoethnography methodology can be an effective tool in analyzing such deceptive designs in VR games. This method allows researchers to actively engage with the game as players, gain first-hand experience with deceptive game designs players' frustrations and excitements, and systematically document their experiences over time~\cite{rapp2018autoethnography}. 
In addition, autoethnography encourages critical self-reflection by requiring researchers to question their interpretations and challenge their assumptions~\cite{rapp2018autoethnography,adams2016handbook}. This introspective process enables nuanced examinations of deceptive game design. The unique combination of subjective player experience and critical researcher analysis makes autoethnography an ideal methodology for answering our research questions.




\section{Methodology}
\label{sec:methodology}
To investigate and contrast the manifestation of computer deceptive game design in commercially available VR games, we used an autoethnography methodology for data collection and a deductive reflexive thematic data analysis. 





\subsection{Data Collection and Documentation}
\label{subsubsec:method-vr-game}



The lead researcher with expertise in deceptive design, games, and VR played and was fully immersed in all three games. Through a game design mechanics analysis and an analysis of player experience from online communities~\cite{hhadan2024ow2}, 
the researcher played the games from a player's perspective and compiled six diary entries and video-recorded 20 hours of gameplay for \OW over six days until no new game design mechanics were observed and no new thoughts and perceptions emerged. Similarly, \Moss was played at length and finished over nine days, totaling eight hours of gameplay and nine diary entries. \BS was played in single-player mode over 10 days with 18 hours of gameplay total, and generated nine diary entries. All VR gameplay sessions were recorded using a Meta Quest3 headset screen-recording function. The 24 systematic diary entries were maintained on Miro\footnote{Miro---the Virtual Workspace for Innovation.~\url{https://miro.com/}} to document the gameplay experience and observations of potentially deceptive designs encountered, complemented by screenshots from video-recorded gameplay sessions. The diary entries and video recordings were then uploaded to Dovetail\footnote{Dovetail---Thematic Analysis Software.\url{https://dovetail.com/}} for thematic analysis. 


\subsection{VR Deceptive Game Design Assessment Guide Development}
\label{subsec:codebook-development}

To ensure consistency in our process of identifying deceptive design in VR games, we developed a \textit{VR Deceptive Game Design Assessment Guide}, inspired by~\citet{gunawan2021comparative}'s deductive codebook method for comparing deceptive design in mobile and web-based applications. We adopted~\citet{gray2024ontology}'s ontology of deceptive designs as our starting point as it is the most recent and comprehensive framework at the time of our study grounded in the foundational literature in the field (e.g.,~\citet{Brignull2010deceptive,gray2018dark,mathur2021makes}). However, the ontology was not tailored to deceptive design in the VR context. Therefore, we augmented our analysis with additional frameworks that focus on VR and games from~\citet{hadan2024deceived},~\citet{king20233d}, and~\citet{zagal2013dark} due to their coverage in VR-specific~\cite{hadan2024deceived} and game-specific~\cite{zagal2013dark,king20233d} deceptive designs, and considerations for player perception in 3D interfaces~\cite{king20233d}. 
To minimize complexity where appropriate, we merged variances of deceptive designs that shared similarities in their definition and characteristics. 
Overall, our \guide contained eight synthesized themes and 71 distinct deceptive design patterns, as summarized in~\autoref{tab:initial-codebook} in the Appendix.

\subsection{Data Analysis}

Given the autoethnography nature of the study, the lead researcher analyzed data using \textit{deductive reflective thematic analysis} method~\cite{clarke2021thematic}. The method preserves the personal and reflexive nature of the research~\cite{rapp2018autoethnography,fassl2023can,turner2022hard} while enabling a ``theoretically-informed exploration of qualitative data''~\cite[p.~260--263]{clarke2021thematic}. Prior studies have extensively used this method to analyze autoethnographic data (e.g.,~\cite{laato2022balancing,turner2022hard,fassl2023can}). To offer transparency in our findings and interpretations~\cite{rapp2018autoethnography}, we provide a \textit{Reflexivity Statement} in~\autoref{sec:app-reflexivity} that 
follows the recommendations for qualitative research~\cite{may2014reflexivity,saldana2021coding}. 

In the early exploratory stage of our research, we reviewed our initial \OW diary entries.
We noticed trends repeatedly centred around the manifestations of deceptive game design and contexts in which they presented and the potential impacts on the game experience. To systematically organize the information hence forward, we subsequently segmented the diary entries into (1) descriptions of the lead researcher's experience, reactions, thoughts, and feelings upon encountering particular game elements, and (2) reflections on possible presence of deceptive game design and its context. An example diary entry for \Moss gameplay is included in~\autoref{sec:app-sample-diary}.

Our main analysis was conducted in 
three stages. In the first stage, the lead researcher reviewed the data based on our \textit{VR Deceptive Game Design Assessment Guide} (see~\autoref{subsec:codebook-development}), with special attention to potential deceptive designs. In the second stage, the diary entries and screenshots from video gameplay recordings were deductively coded based on the \guide. We used an iterative coding process in which the analysis was constantly challenged and refined on the basis of the researcher's critical thoughts and expertise in deceptive design to interpret gameplay experiences. In the third stage, our research team collectively discussed the results to identify any potentially overlooked aspects of the data. 
In this WIP, we discuss the 15 patterns identified in \BS, and five patterns identified in \Moss, and contrast them with the 17 deceptive game design patterns identified in \OW (see~\autoref{tab:brief-codebook}). 


\begin{table*}[!t]
\caption{This table presents deceptive design patterns we identified in \textit{Overwatch2} (\OW)~\cite{hhadan2024ow2}, \BS (BS), and \textit{Moss Book: II} (\Moss). See~\autoref{sec:app-final-codebooks} for specific game mechanics associated with each deceptive pattern and corresponding synthesized diary entries.}
\label{tab:brief-codebook}
\resizebox{2\columnwidth}{!}{%
\begin{tabular}{@{}lllccc@{}}
\toprule
\textbf{\textit{Theme}/Subtheme} & \textbf{Code} & \textbf{Description} & \textbf{OW2} & \textbf{BS} & \textbf{Moss2} \\ \midrule
\multicolumn{2}{l}{\cellcolor{lightgray!50}\textit{\textbf{Theme 1: Sneaking}}} & \cellcolor{lightgray!50} & \cellcolor{lightgray!50} & \cellcolor{lightgray!50} & \cellcolor{lightgray!50} \\
Hiding Information & Reference Pricing~\faGamepad & Include a price-inflated item or a poorer-quality item with same price, make a current price seem ideal. & \faCheckSquareO & - & -\\ \midrule
\multicolumn{2}{l}{\cellcolor{lightgray!50}\textit{\textbf{Theme 2: Obstruction}}} & \cellcolor{lightgray!50} & \cellcolor{lightgray!50} & \cellcolor{lightgray!50} & \cellcolor{lightgray!50} \\
Creating Barriers  & Price Comparison Prevention & Hide information or block copying/pasting, make it difficult to compare prices across vendors. & - &\faCheckSquareO & -\\
& Intermediate Currencies~\faGamepad & Hide the real price behind a virtual currency, make informed purchase decisions difficult. & \faCheckSquareO  & - & - \\  \midrule
\multicolumn{2}{l}{\cellcolor{lightgray!50}\textit{\textbf{Theme 3: Social Engineering}}} &\cellcolor{lightgray!50} &\cellcolor{lightgray!50} &\cellcolor{lightgray!50} & \cellcolor{lightgray!50} \\
 Social Proof & Parasocial Pressure & Exploit users’ trust in celebrities or other entities with inauthentic endorsements to influence decisions. & - & \faCheckSquareO& -\\ 
 Urgency & Countdown Timers~\faGamepad & Create urgency with fake countdown timers. & \faCheckSquareO & - & - \\
 & Limited Time Message~\faGamepad & Create urgency with fake limited-time offers. & - & \faCheckSquareO & - \\
 & Fear of Missing Out (FOMO) & Exploit anxiety about missing exciting or valuable experiences. & \faCheckSquareO & - & - \\
Personalization & Social Obligation~\faGamepad & Create a sense of obligation to avoid letting down friends in the game or to return a favor. & \faCheckSquareO& - & - \\
& Competition~\faGamepad & Foster envy and competitions between players. &\faCheckSquareO &\faCheckSquareO & - \\ \midrule
\multicolumn{2}{l}{\cellcolor{lightgray!50}\textit{\textbf{Theme 4: Interface Interference}}} & \cellcolor{lightgray!50} & \cellcolor{lightgray!50} & \cellcolor{lightgray!50} & \cellcolor{lightgray!50} \\
Manipulating Choice Architecture & False Hierarchy~\faGamepad & Prioritize certain options, make it difficult to compare choices. & \faCheckSquareO & \faCheckSquareO & - \\
& Visual Prominence~\faGamepad & Highlight distracting elements, causing users to forget or lose focus on their initial intent. & - & \faCheckSquareO & -\\ 
& Bundling~\faGamepad & Group products at a ``special'' price and hide individual costs. & \faCheckSquareO & \faCheckSquareO & - \\
Emotional or Sensory Manipulation & Adorable Design~\faGamepad~\faConnectdevelop & Exploit user preference for adorable and appealing design, foster trust in NPCs* and obscure risks. & \faCheckSquareO & \faCheckSquareO & \faCheckSquareO \\
& NPC/Narrative Manipulation~\faGamepad~\faConnectdevelop & Exploit self-satisfaction, sympathy, or reverse psychology to cause unintended actions. & - & - & \faCheckSquareO \\
& Trick Questions & Use confusing wording or double negatives to manipulate users' choices. & - & \faCheckSquareO & - \\
& Hidden Information & Obscure crucial details or disguising them as unimportant. & - & \faCheckSquareO & \faCheckSquareO \\  \midrule
\multicolumn{2}{l}{\cellcolor{lightgray!50}\textit{\textbf{Theme 5: Distortion}}} & \cellcolor{lightgray!50} & \cellcolor{lightgray!50} & \cellcolor{lightgray!50} & \cellcolor{lightgray!50} \\
 & &  & - & - & - \\ \midrule
 \multicolumn{2}{l}{\cellcolor{lightgray!50}\textit{\textbf{Theme 6: Forced Action}}} & \cellcolor{lightgray!50} & \cellcolor{lightgray!50} & \cellcolor{lightgray!50} & \cellcolor{lightgray!50} \\
Forced Communication or Disclosure & Privacy Zuckering & Trick users into thinking it's essential for the service, lead to overshare personal data. & - & \faCheckSquareO & - \\
Gamification & Pay-to-play~\faGamepad & Restricts core functionalities or progress unless players pay to bypass limitations. & \faCheckSquareO & \faCheckSquareO & - \\
& Power Creep~\faGamepad & Diminish obtained item values over time to drive new purchase. & - & \faCheckSquareO & - \\
& Grinding~\faGamepad & Force users into repetitive tasks to access features. & \faCheckSquareO & - & - \\
& Can't Pause or Save~\faGamepad & Force users to continue playing before they can save progress. & - & - & \faCheckSquareO \\
& Playing by Appointment~\faGamepad & Force players to adhere to the game’s schedule. & \faCheckSquareO & - & - \\
Sunk Cost Fallacy & Endowed Value~\faGamepad & Trap users with their previous investment (time, money, effort). & \faCheckSquareO & \faCheckSquareO & - \\
& Endowed Progress~\faGamepad & Exploit users' aversion to incompleted progression. & \faCheckSquareO & - & - \\
& Waste Aversion~\faGamepad & Set small differences between in-game currency and item costs, lead to additional currency purchases.& \faCheckSquareO & - & - \\ \midrule 
\multicolumn{2}{l}{\cellcolor{lightgray!50}\textit{\textbf{Theme 7: Facilitation}}} & \cellcolor{lightgray!50} & \cellcolor{lightgray!50} & \cellcolor{lightgray!50} & \cellcolor{lightgray!50} \\
Infinite Treadmill~\faGamepad & & Continually expand the game (e.g., new levels, new content) so it never ends.  & \faCheckSquareO & - & - \\ \midrule 
\multicolumn{2}{l}{\cellcolor{lightgray!50}\textit{\textbf{Theme 8: Other}}} & \cellcolor{lightgray!50} & \cellcolor{lightgray!50} & \cellcolor{lightgray!50} & \cellcolor{lightgray!50} \\
Complete the Collection~\faGamepad  &  & Exploit users' desires of completing game item collections. & \faCheckSquareO & \faCheckSquareO & \faCheckSquareO \\ 
 \bottomrule
\multicolumn{6}{l}{*NPC = Non-Player Character. ~\faGamepad~denotes deceptive designs mentioned in the game-related literature~\cite{zagal2013dark,king20233d}. 
~\faConnectdevelop~denotes deceptive designs mentioned in VR-related literature~\cite{hadan2024deceived}.}\\
\end{tabular}%
}
\end{table*}

\subsection{Ethical Considerations and Limitations}
Our research followed the research ethics guidelines throughout our processes~\cite{tpcs2022}. Prior to conducting the autoethnographic diary data collection, we have consulted our institution's ethical review board (REB) and ensure that it is not subject to the ethical review process. While we decided to study single-player mode in \BS to avoid involving others without consent, for \OW that unavoidably needs the involvement of other players, we concentrated on the single-player journey and excluded information about other players in our analysis to maintain their privacy.

Our preliminary analysis did not examine deceptive game designs that rely on social relationships due to privacy and ethical considerations regarding other players. 
Although we transparently provided the lead researcher's background and its potential influence on our findings in our autoethnographic study through a reflexivity statement (see~\autoref{sec:app-reflexivity}), we acknowledge that our findings may not represent a broader player population with varied deceptive design literacy and perceptions~\cite{zagal2013dark,geronimo2020UI,luguri2021shining,frommel2022daily}. Our WIP results examined two VR games and a PC game, which may not cover all the deceptive design patterns across the modalities. However, we are confident that the autoethnographic approach allowed us to experience a wide variety of deceptive tactics present in VR games.



\section{Findings}
\label{sec:findings}

This section presents our preliminary findings of deceptive designs in \Moss and \BS, and contrasts their manifestations in VR with the PC game \OW. For clarity, in the following sections, we denote game mechanics in monospace font (e.g.,~\mechanics{Song Gallery}), use \textit{italics} text for deceptive design (e.g.,~\codestyle{Bundling}) and use quoted italics text for the in-game descriptions. We present patterns identified from three games in~\autoref{tab:brief-codebook} and provide descriptions of specific game mechanics from VR games in~\autoref{tab:VR-finalcodebook} and \OW in~\autoref{tab:OW2-finalcodebook} in Appendix with synthesized diary entries.

\begin{figure*}[!t]
    \centering
    \begin{minipage}{0.33\textwidth}
        \centering
        \includegraphics[width=\textwidth]{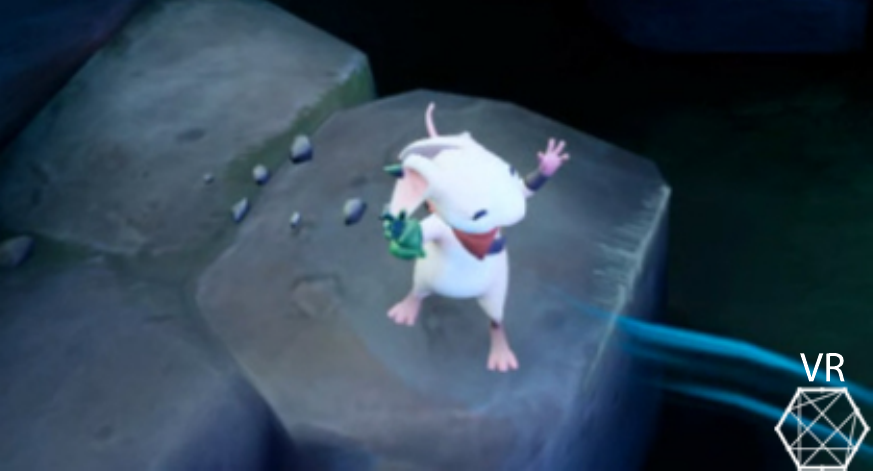}
    \end{minipage} \hfill
    \begin{minipage}{0.33\textwidth}
        \centering
        \includegraphics[width=\textwidth]{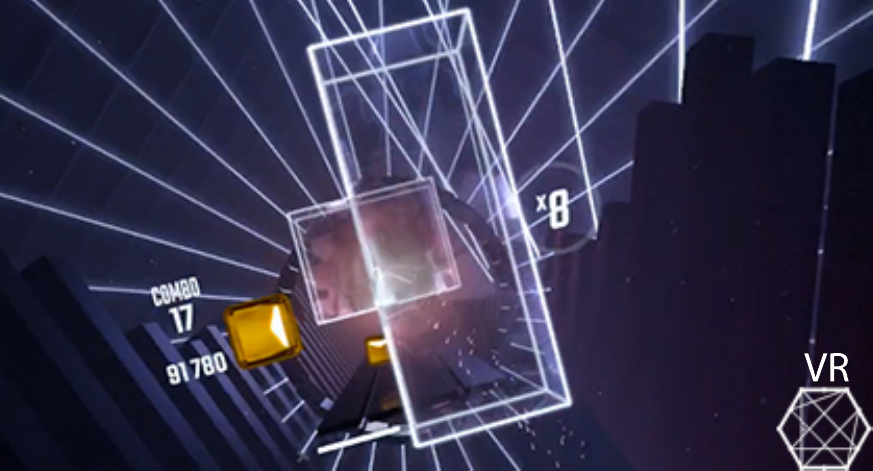}
    \end{minipage} \hfill
    \begin{minipage}{0.33\textwidth}
        \centering
        \includegraphics[width=\textwidth]{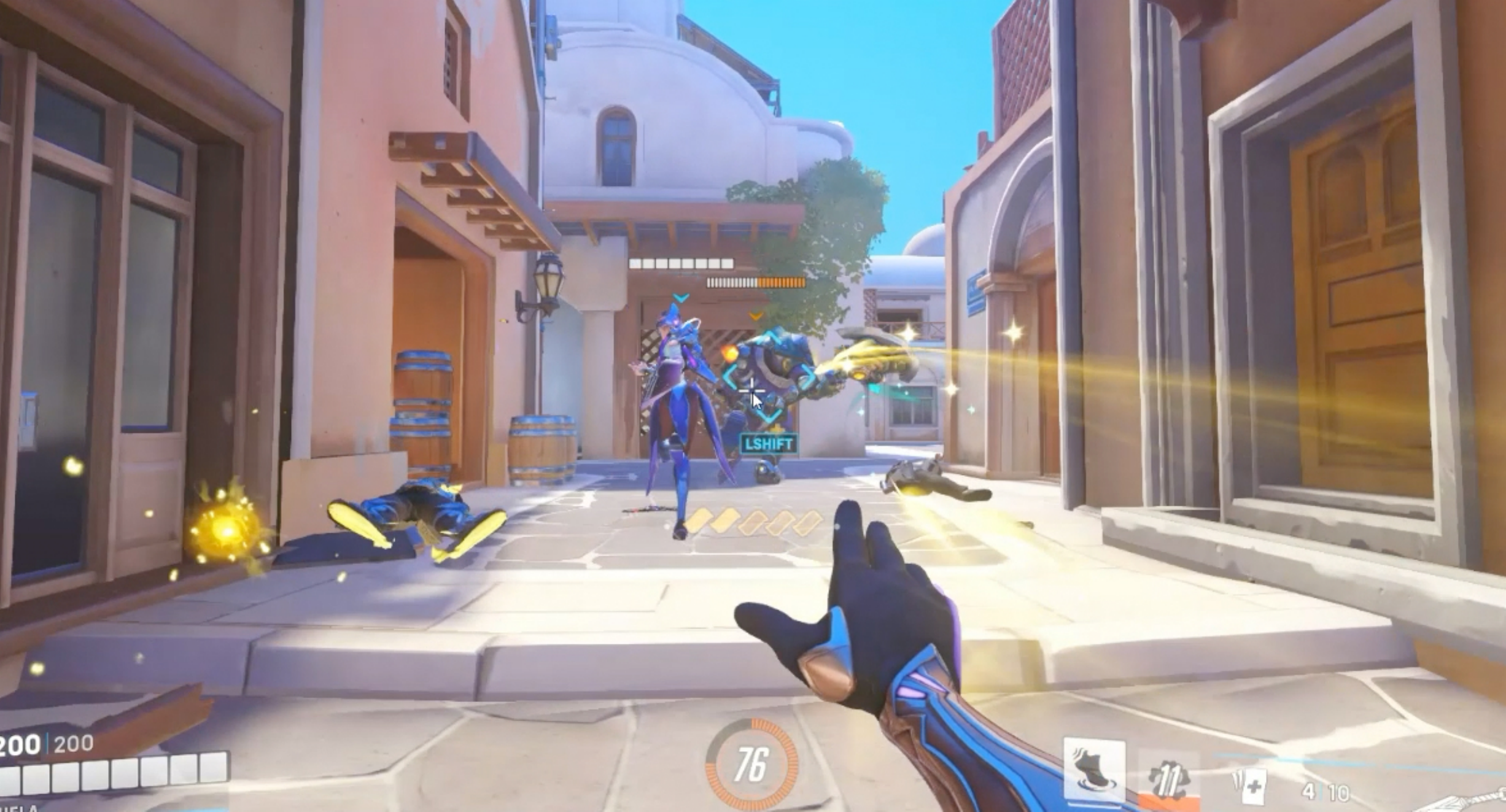}
    \end{minipage}
    \caption{
    Left: sample gameplay moment in \Moss, a VR puzzle adventure game where players guide a tiny mouse character through a fantastical world. Quill, the main character, gives the player a full-body hug, with her tiny arms wrapped tightly around the controller pointers. Center: sample gameplay moment in \BS, a VR rhythm-based block-slashing game, when players are required to physically move left, right or squat down to avoid oncoming obstacles (semi-transparent walls). Right: sample gameplay moment in \OW, a first-person team-based shooter game, when players is using ability to heal (restore) the health of allies at the front by holding down the right-click on mouse.}
    \Description{Left: sample gameplay moment in \Moss, a VR puzzle adventure game where players guide a tiny mouse character through a fantastical world. Quill, the main character, gives the player a full-body hug, with her tiny arms wrapped tightly around the controller pointers. Center: sample gameplay moment in \BS, a VR rhythm-based block-slashing game, when players are required to physically move left, right or squat down to avoid oncoming obstacles (semi-transparent walls). Right: sample gameplay moment in \OW, a first-person team-based shooter game, when players is using ability to heal (restore) the health of allies at the front by holding down the right-click on mouse.}
    \label{fig:screenshots}
\end{figure*}

\subsection{RQ1: What deceptive design patterns are currently implemented in commercial VR games?}

We found that \Moss and \BS shared similarities in deceptive design patterns (see~\autoref{tab:brief-codebook}). Both games used emotional and sensory manipulation (e.g., \codestyle{Adorable Design}) with lovable character designs, eye-catching visuals, and vibrant map environments to create a strong emotional attachment to the game world, potentially leading to longer play sessions and in-game purchases that players may not have otherwise considered. Both games also used \codestyle{Hidden Information} tactics through the fully immersive nature of VR and intense and deeply engaging gameplay. In \BS, the absence of a progress bar could be frustrating for players, as they have to replay the entire song after failing and cannot locate specific parts they fail. The intensity of gameplay could lead to players lose track of time and unintentionally extend the duration of play. In \Moss a progress indicator (\mechanics{Stained Glass Windows}) is subtly integrated into the background map environment. This could cause players to completely miss its existence. As a result, players can feel lost in the engaging storyline and interactive gameplay, and continuously play without a sense of progress. Moreover, \BS and \Moss use \codestyle{Complete the Collection} to fuel players' desire for completion. For instance, in \BS, acquiring all songs in the player's favourite music packs could lead to unintentional overspending of money. Similarly, \Moss has collectible scrolls and character-enhancing armors. While the process of discovering these items 
by repeatedly revisiting completed maps can produce a sense of achievement and a thorough game experience, it could also lead to extensive playtime beyond what is necessary to enjoy the core game.

Beyond game-specific deceptive designs, we found that \BS's \mechanics{Song Purchase System} incorporated \codestyle{Price Comparison Prevention} that inhibits players from comparing the prices of songs and music packs, \codestyle{Limited Time Message} and \codestyle{Bundling} that encourage impulse purchases, \codestyle{Visual Prominence} that attracts players' attention to \textit{``buy music pack''} entirely instead of buying an individual song, and \codestyle{Privacy Zuckering} that requires players' payment information before displaying songs' and music packs' prices. On the other hand, \Moss has largely used deceptive game design patterns that encourage extended playtime and emotional investment from players, including \codestyle{NPC/Narrative Manipulation} that attracts players with its VR-unique interactive elements, such as allowing players to \textit{``pet,''} \textit{``hug,''} \textit{``high-five''} with main characters (see~\autoref{fig:screenshots}) and protect them in battles. While VR technology's limitations prevent a truly tactile experience, the animation of \textit{``hugging''} the player's controller pointers effectively conveys the warmth and gratitude. The main character's sign language communication added another layer of emotional bond and make players feel like they are a part of the adventure and not just outside observers. 
Interestingly, while \Moss incorporated \codestyle{Can't Pause or Save} tactic to keep players continuously play until designated save points, this tactic clashed with the limited VR headset battery life and can frustrate players with significant loss of progress due to a dead battery.
We include the detailed descriptions of patterns identified from VR games, with synthesized diary entries, in Appendix~\autoref{tab:VR-finalcodebook}.

\subsection{RQ2: How do deceptive game designs vary in their manifestations in VR games and PC games?}
\label{subsec:RQ2}


As demonstrated in~\autoref{tab:brief-codebook}, while \OW, \BS, and \Moss all incorporated \codestyle{Adorable Design} and \codestyle{Complete the Collection}, these deceptive design patterns in \OW (e.g., skins with special visual effects, cute weapon decorations) are restricted to the 2D screen. Players interact with them using a mouse, keyboard, or controller, resulting in a clear separation between the player and the game world. On the other hand, \BS and \Moss enable a more dynamic and immersive interaction with VR-unique features. For instance, in \Moss, the VR controllers provide haptic feedback which, coupled with spatial 3D visuals, creates a deeper sense of engagement. The VR environment also allows players to physically lean forward and peer around corners or behind walls to search for strategic paths for the characters. The game also offers a more engaging way of discovering collectibles compared to collecting badges and achievement messages on \OW's 2D interfaces. While \BS places the player in a first-person perspective similar to \OW, unlike the static screen of \OW, players in \BS have a full 360-degree game environment. Players need to not only slash the target blocks but also physically move their body, arms, and legs to avoid virtual obstacles and turn around to reach blocks from different directions (see~\autoref{fig:screenshots}). This full body involvement, combined with the first-person perspective, creates a deeper sense of presence and engaging experience that traditional 2D games cannot easily replicate. 

However, many deceptive design patterns embedded in the game mechanics also have no difference between VR and traditional PC games. For example, both \OW and \BS used \codestyle{Limited Time Message}, \codestyle{Countdown Timer}, \codestyle{Bundling}, and \codestyle{Pay-to-play}, and \codestyle{Competition}. The commonalities are because many deceptive design patterns remain heavily reliant on 2D interfaces even within VR's 3D environments, and therefore, have no significant variations across modalities. 

Lastly, while \BS and \Moss used \codestyle{NPC/Narrative Manipulation}, \codestyle{Trick Questions}, and \codestyle{Power Creep} and \codestyle{Can't Pause or Save} that did not exist in \OW, we believe it is due to differences in game genre rather than the game platform. We include a detailed descriptions of patterns identified from \OW, with synthesized diary entries, in~\autoref{tab:OW2-finalcodebook} in~\autoref{sec:app-final-codebooks}.




\section{Discussion and Next Steps}
\label{sec:discussion}

Although our research is ongoing, we made significant progress in developing a \textit{VR Deceptive Game Design Assessment Guide} tailored to evaluating deceptive design in VR games by synthesizing previous work~\cite{gray2024ontology,king20233d,zagal2013dark,hadan2024deceived}. Our preliminary \guide can serve as a structured foundation for future research and VR design to identify new deceptive practices, analyze player impact, and develop ethical VR game design guidelines. Specifically, researchers can use this \guide to identify new deceptive practices, analyze their impact on players, and develop ethical design guidelines for VR games. VR developers can better understand the different categories of deceptive design and impacts on VR players, create more ethical and player-centric VR experiences, and avoid those causing player frustration, confusion, and pressure in their design. The \guide can also inform regulations or guidelines that promote responsible VR design practices to ensure VR player protection from potentially manipulative tactics in games. 

Our results highlight that the negative effects of some VR deceptive game design patterns like \codestyle{Can't Pause or Save} that force players to engage continuously have only been limited due to hardware constraints of the technology like battery life. VR intensifies player engagement and immersion, and therefore, have the potential to amplify the impacts of deceptive game designs that capitalize on players' emotional and sensory desires (e.g., \codestyle{Adorable Design}, \codestyle{Competition}, and \codestyle{Complete the Collection}). Their existence cautions against the potential for more sophisticated player manipulations in future games. For instance, we can easily foresee that future VR games with \codestyle{Adorable Design} and \codestyle{NPC/Narrative Manipulation} could be extended to NPC conversations that are framed with NPC's inveterate longing~\cite{bogost2010persuasive} to pressure players into in-game purchases (like in~\cite{king20233d}), longer playtime, or guilt-trip players. Such emotional and sensory manipulations can cause greater harm to players compared to those in PC games. Similarly, games like \BS could use \codestyle{Perception Hacking} to trick players restart repetitively by causing unintentional slash on the wrong blocks, or secretly inflating the price of songs in-game with \codestyle{Price Comparison Prevention} by taking advantage that players may not bother to take off headset to look up the price on the website (barrier between virtual and reality~\cite{krauss2024what}). While we found that many deceptive design patterns, such as \codestyle{Pay-to-play} and \codestyle{Countdown Timer}, remain heavily reliant on 2D interfaces, they have the potential to evolve into 3D formats. For instance, \codestyle{Pay-to-play} in VR games could appear as virtual tollbooths at level entrances, while \codestyle{Countdown Timer} might be represented through NPCs' increasingly agitated or aggressive conversations as time runs out. These problems will arise in the near future with more games emerge in VR. Our preliminary findings in this WIP highlight the critical need for ongoing research in deceptive game design in VR to keep up with the rapid evolution of VR technology. 

\textbf{Next Steps:} Building upon our preliminary findings, our first and the immediate next step is to refine and improve the comprehensiveness of the \guide by incorporating a broader scope of literature (e.g.,~\cite{krauss2024what,petrovskaya2021predatory}) and clearer considerations of VR-specific properties (e.g., spatiality, perception, and barriers~\cite{krauss2024what}) that directly influence player deception in VR. Second, our research on deceptive VR game design is still ongoing. As we write this paper, we are expanding our analysis of VR games from diverse genres, which allows us to analyze the VR manifestations of deceptive designs that were not present in our \wip (e.g., \codestyle{Wait To Play}, \codestyle{Grinding}). This will also enable us to contrast and isolate the manifestations of different types of deceptive designs in games from similar genres, or even the same game, across PC and VR, which will result in more meaningful comparisons. Lastly, to complement our autoethnographic methodology, we plan to study VR players' perspectives by analyzing player reviews in online communities and conducting in-depth interviews. We will contrast these player experiences with the findings from our autoethnographic investigation to achieve a comprehensive understanding of how deceptive designs are perceived and experienced. For example, we will closely examine the influence of VR-specific features~\cite{krauss2024what,hadan2024deceived} on player experiences, the perceived severity of various deceptive designs, and how different manifestations shape player perceptions. We hypothesize that certain VR deceptive designs, when combined with other game elements or with different manifestations, may be perceived positively in one game while being viewed as manipulative in another.

In summary, our findings highlighted the critical need for future research in deceptive game designs in VR. Combining findings from our preliminary study and next research steps, our research will enable an understanding of deceptive design manifestations between VR and traditional games, identify how VR's unique features influence the effectiveness of these tactics, and reveal VR-specific deceptions not present in traditional games.
\begin{acks}
This project has been funded by the Office of the Privacy Commissioner of Canada (OPC); the views expressed herein are those of the author(s) and do not necessarily reflect those of the OPC.

L. Nacke also acknowledge support from the Social Sciences and Humanities Research Council (SSHRC) INSIGHT Grant (\#435-2022-0476), Natural Sciences and Engineering Research Council of Canada (NSERC) Discovery Grant (\#RGPIN-2023-03705), and Canada Foundation for Innovation (CFI) John R. Evans Leaders Fund (\#41844) and L. Zhang-Kennedy also acknowledge support from the NSERC Discovery Grant (\#RGPIN-2022-03353)

We also thank the chairs and reviewers for their insightful feedback, which helped us to improve the quality of this manuscript. Thank you to graduate researcher Derrick Wang for his support in resolving technical issues during our paper formatting. Screenshots in this manuscript were from the games and fall under fair use.

\end{acks}

\bibliographystyle{ACM-Reference-Format}
\balance
\bibliography{01-References}

\appendix
\onecolumn
\newpage

\clearpage
\begin{table}[!ht]
\centering
\caption{This table presents our \textit{VR Deceptive Game Design Assessment Guide} and the source literature of themes and codes~\cite{gray2024ontology,zagal2013dark,king20233d,hadan2024deceived}. This \guide served as the foundation of our identification of deceptive designs in the selected VR-based and computer-based games.}
\label{tab:initial-codebook}
\resizebox{0.73\textheight}{!}{%
\begin{sideways}
\begin{tabular}{@{}llll@{}}
\toprule
\textbf{\textit{Theme}/Subtheme} & \textbf{Code} & \textbf{Definition} & \textbf{Correspondence to Taxonomies from Literature} \\ \midrule
\rowcolor{lightgray!50}\multicolumn{2}{l}{\textit{\textbf{Theme 1: Sneaking}}} & &\\
 Bait and Switch& Disguised Ads & ~\deficon~Disguise ads as genuine content, trick users into clicking on it. & ~\coricon~Disguised Ads~\cite{gray2024ontology} \\ \midrule
 Hiding Information & Sneak into Basket & ~\deficon~Sneak items into a user's cart, lead to unintended purchases. & ~\coricon~Sneak into Basket~\cite{gray2024ontology} \\
& Drip Pricing, Hidden Costs, or Partitioned Pricing & ~\deficon~Gradually reveal additional costs, delaying the full price until & ~\coricon~Drip Pricing, Hidden Costs, or Partitioned \\
& & after effort is invested in the purchase process.& Pricing~\cite{gray2024ontology} \\
& Reference Pricing~\faGamepad & ~\deficon~Include a price-inflated item or a poorer-quality item with same & ~\coricon~Reference Pricing~\cite{gray2024ontology},  Cheap Item Placed Next to \\
& & price, making a current price seem like a better deal. & Expensive Item~\cite{king20233d}, Physical Placement of Items~\cite{king20233d}, \\
& & & Anchoring Tricks (Anchoring, Decoy Effect)~\cite{zagal2013dark} \\ \midrule
Perception Hacking~\faConnectdevelop &\multicolumn{1}{c}{-}& ~\deficon~Alter visuals or display out-of-sight ads, causing unwanted & ~\coricon~Perception Hacking~\cite{hadan2024deceived},  Obscuring Reality~\cite{hadan2024deceived}\\
& & and unintentional interaction with objects (e.g., ads)  & \\ \midrule
(De)contextualizing Cues & Conflicting Information & ~\deficon~Confuse users by presenting contradictory details, making & ~\coricon~Conflicting Information~\cite{gray2024ontology} \\
& & them unsure and likely to accept potentially unfavorable defaults.  & \\
 & Information without context & ~\deficon~Hide relevant information or controls, making it difficult for & ~\coricon~Information without context~\cite{gray2024ontology} \\ 
& & users to find what they need.  & \\ \midrule
\rowcolor{lightgray!50}\multicolumn{2}{l}{\textit{\textbf{Theme 2: Obstruction}}} & &\\
Roach Motel& Immortal Accounts & ~\deficon~Trap users by making account deletion difficult or impossible. & ~\coricon~Immortal Accounts~\cite{gray2024ontology} \\
 & Dead Ends & ~\deficon~Frustrate users by hiding information behind broken links or & ~\coricon~Dead Ends~\cite{gray2024ontology} \\
 & & redirects, preventing them from finding desired information. & \\ \midrule
Creating Barriers & Price Comparison Prevention &  ~\deficon~Hide information or block copying/pasting, making it difficult & ~\coricon~Price Comparison Prevention~\cite{gray2024ontology} \\
& & to compare prices across vendors. & \\
 & Intermediate Currencies~\faGamepad & ~\deficon~Hide the real price behind a virtual currency, making informed & ~\coricon~Intermediate Currencies~\cite{gray2024ontology}, Use of Multiple \\
 & & purchase decisions difficult.  & Currencies~\cite{king20233d}, Premium Currency~\cite{zagal2013dark} \\ \midrule
Adding Steps  & Privacy Mazes & ~\deficon~Bury privacy controls under layers of confusing menus, &  ~\coricon~Privacy Mazes~\cite{gray2024ontology}\\
& & hindering users from making informed choices.  & \\ \midrule
Reality Distortion~\faConnectdevelop &\multicolumn{1}{c}{-}& ~\deficon~Conceal parts of reality with digital objects, through & ~\coricon~Reality Distortion~\cite{hadan2024deceived}, Obscuring Reality~\cite{hadan2024deceived}, \\
& & daily wear and photorealistic graphics, leading to false perception & Persistent Exposure to Manipulation~\cite{hadan2024deceived}\\ 
& &  and change of behaviours. & \\  \midrule
\rowcolor{lightgray!50}\multicolumn{2}{l}{\textit{\textbf{Theme 3: Social Engineering}}} & &\\
Scarcity or Popularity Claims & High Demand~\faGamepad & ~\deficon~Create a false sense of urgency by exaggerating product  & ~\coricon~Anchoring Tricks (item popularity)~\cite{zagal2013dark} \\
&  & popularity, tricking users into rushed purchases. & ~\coricon~High Demand~\cite{gray2024ontology}\\ \midrule
Social Proof & Low Stock~\faGamepad & ~\deficon~Pressure users into impulsive purchases by falsely indicating & ~\coricon~Low Stock~\cite{gray2024ontology},Artificial Scarcity (only a certain \\
& & limited product availability. & number of items available)~\cite{zagal2013dark}\\
 & Endorsements and Testimonials & ~\deficon~Portray biased or fabricated testimonials as genuine to & ~\coricon~Endorsements and Testimonials~\cite{gray2024ontology}, Friend Spam /\\
 & & influence users' purchase decisions.  &  Impersonation (friends follow your endorsement)~\cite{zagal2013dark} \\
 & Parasocial Pressure~\faGamepad~\faConnectdevelop & ~\deficon~Exploit users' trust in celebrities or other entities with &  ~\coricon~Parasocial Pressure~\cite{gray2024ontology}, Anchoring Tricks (friends' \\
 & & inauthentic endorsements to influence users' decisions. & purchase activities)~\cite{zagal2013dark}, NPC Asks or Suggests a \\
 & & & Purchase~\cite{king20233d}, Hyperpersonalization (recreations of \\
 & & & their trusted people)~\cite{hadan2024deceived}, Instructed/Told \\
 & & & to Do~\cite{king20233d} \\ \midrule
Urgency & Activity Messages & ~\deficon~Create urgency with fake activity notifications of others. & ~\coricon~Activity Messages~\cite{gray2024ontology} \\
 & Countdown Timers~\faGamepad & ~\deficon~create urgency with fake countdown timers. & ~\coricon~Countdown Timers~\cite{gray2024ontology}, Artificial Scarcity~\cite{zagal2013dark} \\
 & Limited Time Messages~\faGamepad & ~\deficon~Create urgency with fake limited-time offers. & ~\coricon~Limited Time Messages~\cite{gray2024ontology}, Artificial Scarcity~\cite{zagal2013dark}\\
 & Fear of Missing Out (FOMO)~\faGamepad & ~\deficon~Exploit anxiety about missing exciting or valuable experiences. & ~\coricon~Fear of Missing Out~\cite{zagal2013dark,king20233d}\\ \midrule
Personalization &Confirmshaming & ~\deficon~Guilt-trip users with manipulative language.& ~\coricon~Confirmshaming~\cite{gray2024ontology}\\
& Social Obligation~\faGamepad & ~\deficon~Create a sense of obligation to avoid letting down friends in the & ~\coricon~Aesthetic Manipulations (feel bad for not helping \\
& & game or to return a favor. & team)~\cite{zagal2013dark}, Social Obligation / Guilds~\cite{zagal2013dark}, Players \\
& & & Feels Ashamed/Guilty~\cite{king20233d}, Reciprocity~\cite{zagal2013dark}\\
& Competition~\faGamepad & ~\deficon~Foster envy and competitions between players. & ~\coricon~Competition~\cite{zagal2013dark} \\ \midrule
Encourage Anti-Social Behavior~\faGamepad &\multicolumn{1}{c}{-}&  ~\deficon~Incentivize users' dishonest or harmful actions to gain benefits. & ~\coricon~Encourages Anti-Social Behavior~\cite{zagal2013dark}\\
\bottomrule
\multicolumn{4}{l}{\textit{Note.} ~\faGamepad~denotes deceptive designs mentioned in the game-related literature~\cite{zagal2013dark,king20233d}. 
~\faConnectdevelop~denotes deceptive designs mentioned in VR-related literature~\cite{hadan2024deceived}.}\\
\end{tabular}%
\end{sideways}
}
\end{table}
\clearpage
\begin{table}[!ht]
\centering
\caption*{Table 2 Continued. This table presents our \textit{VR Deceptive Game Design Assessment Guide} and the source literature of themes and codes~\cite{gray2024ontology,zagal2013dark,king20233d,hadan2024deceived}. This \guide served as the foundation of our identification of deceptive designs in the selected VR-based and computer-based games.}
\label{tab:initial-codebook-part2}
\resizebox{0.73\textheight}{!}{%
\begin{sideways}
\begin{tabular}{@{}llll@{}}
\toprule
\textbf{\textit{Theme}/Subtheme} & \textbf{Code} & \textbf{Definition} & \textbf{Correspondence to Taxonomies from Literature} \\ \midrule
\rowcolor{lightgray!50}\multicolumn{2}{l}{\textit{\textbf{Theme 4: Interface Interference}}} & &\\
Manipulating Choice Architecture & False Hierarchy~\faGamepad &  ~\deficon~Prioritize certain options, making it difficult to compare choices & ~\coricon~False Hierarchy~\cite{gray2024ontology}, Anchoring Tricks (False \\
& &  and potentially leading to unintended selections. & Hierarchy)~\cite{zagal2013dark}, Aesthetic Manipulations (bigger \\
& & & checkbox)~\cite{zagal2013dark}, ``Premium Value'' or ``Rare'' Items~\cite{king20233d}  \\
 & Visual Prominence~\faGamepad &  ~\deficon~Highlight distracting elements, causing users to forget or lose & ~\coricon~Visual Prominence~\cite{gray2024ontology}, Exciting Sound Effect~\cite{king20233d}, \\
 & &  focus on their initial intent. & Aesthetic Manipulations (red  ``yes'' button)~\cite{zagal2013dark},\\
 & & &  UI Colours/Animation~\cite{king20233d} \\
 & Bundling~\faGamepad & ~\deficon~Group products at a ``special'' price, potentially hiding individual & ~\coricon~Anchoring Tricks (Discount Bundles)~\cite{zagal2013dark} \\
 & &  costs and leading to uninformed purchases. & Bundling~\cite{gray2024ontology}, Bulk-buy Discount~\cite{king20233d} \\
 & Pressured Selling & ~\deficon~Highlight or pre-select expensive options, potentially leading & ~\coricon~Pressured Selling~\cite{gray2024ontology} \\
 & &  users to overlook more affordable choices.&\\ \midrule
Bad Defaults~\faGamepad &\multicolumn{1}{c}{-}& ~\deficon~Set default options that benefit the company, forcing users to & ~\coricon~Bad Defaults~\cite{gray2024ontology}, Accidental Purchase (expensive \\ 
& & manually change settings to avoid privacy risks. & ones selected by default)~\cite{zagal2013dark}, Aesthetic Manipulations \\
& & & (checkbox opt-in by default)~\cite{zagal2013dark} \\ \midrule
Misclick~\faGamepad~\faConnectdevelop &\multicolumn{1}{c}{-}& ~\deficon~Exploit strategic button placement and absence of confirmation & ~\coricon~Obscuring Reality~\cite{hadan2024deceived}, Perception Hacking (offset\\
& & screen, leading to accidental actions. &  visual cue in cursor-jacking)~\cite{hadan2024deceived}, Accidental Purchase \\
& & &  (accidental stray tap of the screen)~\cite{zagal2013dark}\\ \midrule 
Emotional or Sensory Manipulation & Adorable design~\faGamepad~\faConnectdevelop & ~\deficon~Exploit user preference for adorable design to make the game & ~\coricon~Cuteness~\cite{gray2024ontology}, NPC is cute or very likeable~\cite{king20233d}, \\
& &  appealing, fostering trust in characters, and obscuring risks. & Hyperpersonalization (content tailored to preference\\
& & & of an individual)~\cite{hadan2024deceived}, Aesthetic Manipulations\\
& & &  (Subconscious Associations)~\cite{zagal2013dark} \\
 & Positive or Negative Framing~\faGamepad & ~\deficon~Hide or downplaying crucial information through visual cues, & ~\coricon~Positive or Negative Framing~\cite{gray2024ontology}, Users Images\\
 & & leading to biased or distracted decisions. &  Associated with Wealth~\cite{king20233d}, Aesthetic Manipulations \\
 & & &  (Toy with Emotions)~\cite{zagal2013dark}\\
 & NPC/Narrative Manipulation*~\faGamepad~\faConnectdevelop & ~\deficon~Exploit self-satisfaction, sympathy, or reverse psychology to & ~\coricon~Players Feel Good for Helping NPC~\cite{king20233d}, Sympathy/\\
 & & cause unintended actions. & pity for NPC~\cite{king20233d}, Want to Prove NPC Wrong (to \\
 & & & counter NPC's taunt)~\cite{king20233d}, Hyperpersonalization \\
 & & & (idealized interaction partners)~\cite{hadan2024deceived}, Artificial \\
 & & &  Prosthetic Memory/Empathy-Based Manipulation~\cite{hadan2024deceived}\\
 & General Emotional Manipulation~\faGamepad & ~\deficon~Non-specific descriptions of emotional manipulation. & ~\coricon~General Emotional Manipulation~\cite{king20233d}\\ \midrule
Trick Questions &\multicolumn{1}{c}{-}& ~\deficon~Use confusing wording or double negatives to manipulate users' & ~\coricon~Trick Questions~\cite{gray2024ontology} \\
& &  choices. & \\ \midrule
Choice Overload &\multicolumn{1}{c}{-}& ~\deficon~Overwhelm users with too many options, hindering comparison & ~\coricon~Choice Overload~\cite{gray2024ontology} \\ 
& & and potentially causing them to miss important information. & \\
Hidden Information &\multicolumn{1}{c}{-}& ~\deficon~Obscure crucial details or disguising them as unimportant. & ~\coricon~Hidden Information~\cite{gray2024ontology} \\ \midrule
Language Inaccessibility & Wrong Language & ~\deficon~Display crucial information in a foreign language, making it & ~\coricon~Wrong Language~\cite{gray2024ontology} \\
& &  inaccessible to users. & \\
 & Complex Language & ~\deficon~Use hard-to-understand words and sentence structures, & ~\coricon~Complex Language~\cite{gray2024ontology} \\
 & & hindering users' informed decisions. & \\ \midrule
Feedforward Ambiguity &\multicolumn{1}{c}{-}& ~\deficon~Create a gap between users' predicted result from the available & ~\coricon~Feedforward Ambiguity~\cite{gray2024ontology} \\
& & information and the actual outcome of their actions. & \\ \midrule
\rowcolor{lightgray!50}\multicolumn{2}{l}{\textit{\textbf{Theme 5: Distortion}}} & &\\
Cognitive Bias & Illusion of Mastery~\faGamepad & ~\deficon~Deceive players about their skill level, making them feel better & ~\coricon~Illusion of Control (e.g., manipulated matchups,\\
& & than they actually are and encouraging more gameplay. &  undisclosed random win chance)~\cite{zagal2013dark}\\
& Optimism and Frequency Biases~\faGamepad & ~\deficon~Exploit optimism bias, frequency bias, and clustering illusion to &~\coricon~Optimism and Frequency Biases~\cite{zagal2013dark} \\
& & make players feel luckier and skilled, encouraging continued play. & \\
 & False Memory Implantation~\faConnectdevelop & ~\deficon~Exploit users' memory flaw and source confusing to induce false & ~\coricon~False Memory Implantation~\cite{hadan2024deceived} \\
 & &  memory and influence their behaviour. & \\
 & The illusion of objectivity~\faConnectdevelop & ~\deficon~First-person experiences inherently biased by creator influence, & ~\coricon~The illusion of objectivity in XR experience~\cite{hadan2024deceived}\\
 & &  creating a false sense of objectivity. &\\
\bottomrule
\multicolumn{4}{l}{\textit{Note}. *NPC = Non-Player Character. ~\faGamepad~denotes deceptive designs mentioned in the game-related literature~\cite{zagal2013dark,king20233d}. 
~\faConnectdevelop~denotes deceptive designs mentioned in VR-related literature~\cite{hadan2024deceived}.}\\
\end{tabular}%
\end{sideways}
}
\end{table}
\clearpage
\begin{table}[!ht]
\centering
\caption*{Table 2 Continued. This table presents our \textit{VR Deceptive Game Design Assessment Guide} and the source literature of themes and codes~\cite{gray2024ontology,zagal2013dark,king20233d,hadan2024deceived}. This \guide served as the foundation of our identification of deceptive designs in the selected VR-based and computer-based games.}
\label{tab:initial-codebook-part3}
\resizebox{0.73\textheight}{!}{%
\begin{sideways}
\begin{tabular}{@{}llll@{}}
\toprule
\textbf{\textit{Theme}/Subtheme} & \textbf{Code} & \textbf{Definition} & \textbf{Correspondence to Taxonomies from Literature} \\ \midrule
\rowcolor{lightgray!50}\multicolumn{2}{l}{\textit{\textbf{Theme 6: Forced Action}}} & &\\
Nagging~\faGamepad &\multicolumn{1}{c}{-}& ~\deficon~Disrupt user focus with repeatedly unwanted interruptions to & ~\coricon~Nagging~\cite{gray2024ontology}, Badges / Endowed Progress (red dot  \\
& & push actions they would rather avoid. & reminder of left off progress)~\cite{zagal2013dark}, Prompted to Buy \\
& & & Robux~\cite{king20233d}, Aesthetic Manipulations (Over-the-top  \\
& & & Feedback)~\cite{zagal2013dark}\\ \midrule
Forced Continuity &\multicolumn{1}{c}{-}& ~\deficon~Continuity traps users in unwanted subscriptions by making & ~\coricon~Forced Continuity~\cite{gray2024ontology}  \\
& & cancellation difficult and auto-renewal opaque. &\\ \midrule
Forced Registration &\multicolumn{1}{c}{-}& ~\deficon~Require account creation for tasks that shouldn't need it, & ~\coricon~Forced Registration~\cite{gray2024ontology} \\
& & potentially extracting unnecessary personal data. & \\ \midrule
Forced Communication or Disclosure & Privacy Zuckering & ~\deficon~Trick users into thinking it's essential for the service, & ~\coricon~Privacy Zuckering~\cite{gray2024ontology} \\
& & leading to overshare personal data.  & \\
 & Friend Spam~\faGamepad~\faConnectdevelop & ~\deficon~Exploit social connections by sending spam to users' contact & ~\coricon~Friend Spam~\cite{gray2024ontology}, Friend Spam/Impersonation~\cite{zagal2013dark}, \\
 & & lists and resulting in unwanted contact to other users. & Hyperpersonalization (recreate trusted people)~\cite{hadan2024deceived} \\
 & Address Book Leeching & ~\deficon~Steal users' contact information under their false impression &Address Book Leeching~\cite{gray2024ontology}\\
 & & that only vital information will be collected. & \\
 & Social Pyramid~\faGamepad  & ~\deficon~Push users into recruiting others by tying it to desired features. & ~\coricon~Social Pyramid~\cite{gray2024ontology}, Social Pyramid Scheme~\cite{zagal2013dark} \\ \midrule
Gamification & Pay-to-Play~\faGamepad & ~\deficon~Initially advertised as free, later restricts core functionalities or &  ~\coricon~Pay-to-Play~\cite{gray2024ontology}, Pay Wall~\cite{zagal2013dark}, Pay to Skip~\cite{zagal2013dark}, \\
& & progress unless players pay to bypass limitations (limited actions, & Recurring Fee~\cite{zagal2013dark}, Real Money Spend Expected~\cite{king20233d}, \\
& &  timers, ads). & Purchase Feels Part of Narrative~\cite{king20233d}\\
& Pay to Win~\faGamepad & ~\deficon~Real-money purchases grant significant in-game advantages. & ~\coricon~Pay to Win~\cite{zagal2013dark}\\
& Power Creep~\faGamepad & ~\deficon~Diminish obtained item values over time to drive new purchase. & ~\coricon~Power Creep~\cite{zagal2013dark}\\
 & Grinding~\faGamepad & ~\deficon~Force users into repetitive tasks to access features. & ~\coricon~Grinding~\cite{gray2024ontology}, Grinding~\cite{zagal2013dark} \\
 & Random Rewards~\faGamepad & ~\deficon~Encourage users to spend money on chance-based rewards. & ~\coricon~Variable Rewards~\cite{zagal2013dark}, Gambling/LootBoxes~\cite{zagal2013dark}, \\
 & & & Gacha/lootbox/random~\cite{king20233d}, ``Gambling''~\cite{king20233d}\\
 & Can't Pause or Save~\faGamepad & ~\deficon~Force users to continue playing before they can save progress. & ~\coricon~Can't Pause or Save~\cite{zagal2013dark}\\
 & Daily Challenges~\faGamepad & ~\deficon~Encourage players to revisit the game daily.& ~\coricon~Daily Challenges~\cite{zagal2013dark} \\
 & Playing by Appointment~\faGamepad &  ~\deficon~Force players to adhere to the game's schedule.& ~\coricon~Playing by Appointment~\cite{zagal2013dark}\\
 & Wait To Play~\faGamepad & ~\deficon~Impose arbitrary wait times on players through in-game timers. & ~\coricon~Wait To Play~\cite{zagal2013dark}\\ \midrule
Sunk Cost Fallacy & Endowed Value ~\faGamepad & ~\deficon~Trap users with their previous investment (time, money, effort). & ~\coricon~Invested / Endowed Value~\cite{zagal2013dark}\\
 & Endowed Progress~\faGamepad & ~\deficon~Exploit users' aversion to incompleted progression. & ~\coricon~Badges / Endowed Progress~\cite{zagal2013dark}\\
 & Waste Aversion~\faGamepad & ~\deficon~Set small differences between in-game currency and item costs, & ~\coricon~Must Buy More Robux (currency) Than \\
 & & prompting additional currency purchases. & Needed~\cite{king20233d}, Waste Aversion~\cite{zagal2013dark}\\ \midrule
Attention Capture & Auto-Play & ~\deficon~Auto-start videos, causing excessive or unwanted views.& ~\coricon~Auto-Play~\cite{gray2024ontology} \\
 & Interest Exploitation~\faConnectdevelop & ~\deficon~Leverage users' interests to capture their attention. & ~\coricon~Hyperpersonalization (idealized partners)~\cite{hadan2024deceived}\\ \midrule
\rowcolor{lightgray!50}\multicolumn{2}{l}{\textit{\textbf{Theme 7: Facilitation}}} & &\\
Low Barrier & Low Barrier Purchase~\faGamepad & ~\deficon~Minimize steps, and complexity during spending actions, & ~\coricon~Purchasing Is Too Easy~\cite{king20233d}, \\
& &  encouraging impulsive purchases. & ~\coricon~Purchase Prompt is Immediate~\cite{king20233d}\\ \midrule
Infinite Treadmill~\faGamepad &\multicolumn{1}{c}{-}& ~\deficon~Continually expanding the game (e.g., new levels, new content). & ~\coricon~Infinite Treadmill~\cite{zagal2013dark}\\ \midrule
\rowcolor{lightgray!50}\multicolumn{2}{l}{\textit{\textbf{Theme 8: Other}}} & &\\
Complete the Collection~\faGamepad &\multicolumn{1}{c}{-}&  ~\deficon~Exploit users' desires of completing game item collections. & ~\coricon~Complete the Collection~\cite{zagal2013dark}\\ \midrule
Data Fueled Manipulation~\faConnectdevelop &\multicolumn{1}{c}{-}&  ~\deficon~Capture extensive user data (movement, behavior, surroundings) & ~\coricon~Data Fuels Privacy Risks and Manipulation~\cite{hadan2024deceived}, \\
& &  enabling potential deanonymization and tailored manipulation.& Undesired Access and Data Use~\cite{hadan2024deceived}, Insecurity \\
& & & Worsen Manipulation~\cite{hadan2024deceived}\\ \midrule
Blurry Boundary~\faConnectdevelop &\multicolumn{1}{c}{-}& ~\deficon~Photorealistic visual designs blur the lines between reality and & ~\coricon~Extreme Realism blurs the boundary between  \\ 
& &  ads, potentially misleading users about product quality.& virtual and reality~\cite{hadan2024deceived}, Obscuring Reality~\cite{hadan2024deceived}, \\
& & & Persistent Exposure to Manipulation~\cite{hadan2024deceived}\\
\bottomrule
\multicolumn{4}{l}{\textit{Note}. ~\faGamepad~denotes deceptive designs mentioned in the game-related literature~\cite{zagal2013dark,king20233d}. 
~\faConnectdevelop~denotes deceptive designs mentioned in VR-related literature~\cite{hadan2024deceived}.}\\
\end{tabular}%
\end{sideways}
}
\end{table}

\clearpage  
\newpage
\section{Sample Diary Entry Structure}
\label{sec:app-sample-diary}

\begin{table}[!ht]
\centering
\caption{Structural outline of a sample diary entry from a \Moss gameplay session on February 1, 2024. The diary was documented in Miro given its feasibility in linking videos, screenshots, and diary texts. From left to right: (1) description of the lead researcher's experience, reactions, thoughts, and feelings complemented by corresponding screenshots of game design elements, and (2) post-gameplay reflection on possible presence of deceptive game design and its context.}
\label{tab:diary-mockup}
\resizebox{0.9\textwidth}{!}{%
\begin{tabular}{@{}p{0.5\columnwidth}|p{0.5\columnwidth}@{}}
\toprule
\multicolumn{2}{l}{\textbf{Date:} February 1, 2024}   \\ 
\multicolumn{2}{l}{\textbf{Estimated Time:} $\sim$3.5 hours including diary writing} \\ \midrule
\multicolumn{2}{l}{\textbf{{[}Imported Video Gameplay Recording{]}} } \\ \midrule
\textbf{{[}\textit{Diary}{]}} & \textbf{{[}Reflection Note:{]}}\\
\textit{Today's session with Moss Book II produced a strong emotional moment that further made me more emotionally invested in Quill, the brave mouse protagonist. She had to fight a scary owl (a winged tyrant) that threw her and her uncle off a cliff. I was on the edge of my seat as Quill went after her uncle. The game then showed a really touching scene where Quill found her uncle Argus died. The moment that she holds Argus' body and looks at my direction for help made me feel like I was right there with Quill on her suffering.}& I really love how Quill interacts with me directly - it makes the game so much more engaging. Her reactions to my moves are spot on and she reacts to my every move and gives me a high-five when I do something right. Even when I mess up, she's still there showing she cares about me and bears with me while I try to solve the puzzles. It's like she's a real character to me, not just some avatar. This feeling motivates me to keep going and solve all the puzzles and guide her through the story.\\
\multicolumn{1}{r|}{ \textbf{{[}Screenshots{]}} }& \\
\textit{As I and Sahima found Quill under the cliff, she surprised me with a hug. While the limitations of VR technology prevented a fully tactile experience, the animation of ``hugging'' my controller pointers with her whole body effectively conveyed the warmth and gratitude in her embrace. This simple gesture touched deeply and reminded me of the bond we had built throughout our shared adventure. It was no longer just about guiding Quill through puzzles anymore; it was about protecting a close friend.} & The emotional weight of the hug was further amplified when we found what happened to Quill's uncle. His death made the story of Moss even more meaningful and gave me a stronger sense of purpose. I became even more determined to keep Quill safe and avenge her uncle's loss because I knew she was counting only on me now.\\
\multicolumn{1}{r|}{ \textbf{{[}Screenshots{]}} } & \textbf{{[}Next Session:{]}}\\
\textit{The way the game implemented it for Quill's hug was cool, but imagine if we could feel every detail – like the warmth of her fur or the gentle squeeze of her tiny paws. That would be so immersive and emotional. I bet it would change the way we feel about the characters we're playing with.}& In my next session, I'll be looking for other methods that the game uses to enhance player immersion and emotional connection.  It will be interesting to see how Moss Book II continues to use storytelling and VR technology to create a truly impactful experience.\\
 \bottomrule
\end{tabular}%
}
\end{table}

\twocolumn
\clearpage
\newpage
\section{Reflexivity Statement}
\label{sec:app-reflexivity}

To ensure the rigour of our qualitative research, we follow the recommendations by~\citet{may2014reflexivity} and~\citet{saldana2021coding} by providing a \textit{Reflexivity Statement}. In this statement, we acknowledge the potential biases that may arise during the data coding and theme development process due to the lead researcher's cultural, education, and gaming background. The reflexive thematic analysis was performed by the lead researcher, who has a strong research background in deceptive design and has previously published on the topic in both Extended Reality (including VR) and traditional PC-based games~\cite{hadan2024deceived,hhadan2024ow2,hadan2024pop}. The lead researcher also has several years of experience researching privacy and security issues from the user's perspective. This interdisciplinary background and experience align well with this reflexive method, as it identifying deceptive game designs and potential concerns from players' perspectives. In addition, our research team also offers diverse expertise such as interaction design, user experience, privacy and cybersecurity, and games user research. This collective expertise strengthens the comprehensiveness and quality of our research.

Our autoethnography methodology puts the lead researcher into a player perspective. While the lead researcher's expertise and extensive experience in deceptive design ensured a rich identification of these patterns in the game, we still note that her personality might have influenced the focus and the interpretation of autoethnographic data. The lead researcher, as a player, identified herself as empathetic and sensitive and tend to be caring and compassionate. This personality allowed for picking up on subtle emotional manipulation and patterns that exploit players' attachment to in-game characters. In addition, the lead researcher has a strong focus on observable facts and more straightforward outcomes based on concrete evidence. While this personality might have limited the lead researcher's interpretation of hypothetical future consequences of deceptive design, her expertise enabled a thorough understanding of player manipulation currently in the games.  

We also note that the lead researcher's personal experience with video games has influenced this research. The lead researcher was introduced to video games at a young age. This early experience includes a wide range of North American and Asian titles across different genres, played on both computer and PlayStation platforms. In 2017, the lead researcher was introduced to \textit{Overwatch}, which enabled a reunion between her family members during the COVID-19 pandemic, until Blizzard shut down the server in China in 2023~\cite{ign_2024}. To date, the lead researcher has played \textit{Overwatch} (both \textit{OW1} and \textit{OW2}) for about 7 years, more than 3,000 hours of gameplay in different game modes and characters. This experience, combined with the lead researcher's extensive exposure and experience with other games, fosters a deep understanding of player perspectives and various game design patterns.

In summary, the lead researcher's background presents a valuable resource to our autoethnographic research, as it allows us to have a nuanced documentation and analysis of potential deceptive game elements. While we acknowledge that the lead researcher's personality and personal connection to \OW could influence our research findings, we see this as a strength. This deep familiarity with the game equips the lead researcher with a nuanced understanding of its strengths and weaknesses, which enabled a more comprehensive analysis.


\onecolumn
\newpage
\section{Deceptive Game Design Patterns Identified from Overwatch2 and VR Games}
\label{sec:app-final-codebooks}

\begin{table*}[!ht]
\centering
\vspace{-3mm}
\caption{Using our \textit{VR Deceptive Game Design Assessment Guide} in~\autoref{tab:initial-codebook}, this table presents deceptive design patterns we identified within Overwatch 2 (OW2) based on our previous research on this game~\cite{hhadan2024ow2}.}
\vspace{-2mm}
\label{tab:OW2-finalcodebook}
\resizebox{0.95\textwidth}{!}{%
\begin{tabular}{@{}llll@{}}
\toprule
\textbf{\textit{Theme}/Subtheme} & \textbf{Code} & \textbf{OW2 Game Mechanics} & \textbf{Synthesized Diary Entries$^a$} \\ \midrule
\multicolumn{2}{l}{\cellcolor{lightgray!50}\textit{\textbf{Theme 1: Sneaking}}} & \cellcolor{lightgray!50} & \cellcolor{lightgray!50} \\
Hiding Information & Reference Pricing~\faGamepad & In-Game Shop &~\deficon~I feel discounted/cheaper items are better deals compared to \\
& & &  full-priced/expensive ones displayed alongside. \\  \midrule
\multicolumn{2}{l}{\cellcolor{lightgray!50}\textit{\textbf{Theme 2: Obstruction}}} & \cellcolor{lightgray!50} & \cellcolor{lightgray!50} \\
Creating Barriers  & Intermediate Currencies~\faGamepad & OW Coins &~\deficon~ I feel overwhelmed by five distinct currencies (OW Coins,\\
& & &  Legacy Credits, Competitive Coins, Overwatch League Tokens, \\
& & &  Mythic Prisms), each has different exchange rates to real money \\ 
& & &  and restricted to specific item purchases.\\ \midrule
\multicolumn{2}{l}{\cellcolor{lightgray!50}\textit{\textbf{Theme 3: Social Engineering}}} & \cellcolor{lightgray!50} & \cellcolor{lightgray!50} \\
Urgency  & Countdown Timers~\faGamepad & Battle Pass & ~\deficon~ I feel pressured to complete the Battle Pass due to a countdown\\
& & &   timer indicating its end at each season's close. \\
 & & Daily/Weekly/Seasonal Challenges & ~\deficon~ I feel pressured to complete challenges as they refresh daily, \\
 & & &  weekly, and seasonally, indicated by a countdown timer.\\
 & & In-Game Shop & ~\deficon~ I feel pressured to buy bundles and discounted items as they\\
 & & &  refresh weekly, indicated by a countdown timer.\\ 
 & Fear of Missing Out (FOMO)~\faGamepad & Battle Pass&~\deficon~ I fear not completing the Battle Pass before the season ends, as \\
 & & & it means losing the chance to earn its rewards and potentially\\ 
 & & &  wasting the premium fee. \\
 & & Daily/Weekly/Seasonal Challenges & ~\deficon~ I fear not completing the challenges within the time limit, as it \\
 & & & means losing the chance to earn the XPs and currencies.\\
 & & New Character & ~\deficon~ I fear not unlocking the new character from the Battle Pass this\\
 & & &  season, as it means missing the easiest path and quick access.\\
 Personalization & Social Obligation~\faGamepad & New Character & ~\deficon~ I feel obligated to quickly unlock the new character to help\\
 & & & my team in games.\\ 
 & Competition~\faGamepad & Competitive Ranks &~\deficon~I feel jealous of higher-ranked players, which drives me to play \\
 & & & competitive games repeatedly to practice and improve my rank. \\\midrule
\multicolumn{2}{l}{\cellcolor{lightgray!50}\textit{Theme 4: Interface Interference}} & \cellcolor{lightgray!50} & \cellcolor{lightgray!50} \\
Manipulating Choice Architecture & False Hierarchy~\faGamepad & Battle Pass & ~\deficon~I feel buying premium is worthwhile, as nearly all rewards are\\
& & & premium-only.\\
& & Character Item Gallery &~\deficon~ I feel more easily find and click on legendary and epic items, as \\
& & &  they are always at the top of the gallery. \\
 & Bundling~\faGamepad & Battle Pass & ~\deficon~ I feel buying a Battle Pass with 80+ rewards cheaper than \\
 & & & buying them individually, even they are items I don't want.\\
 & & In-Game Shop & ~\deficon~ I feel buying bundled items at a discounted price is cheaper \\
 & & & than buying them individually, even there are items I don't want.\\
Emotional or Sensory Manipulation & Adorable design~\faGamepad~\faConnectdevelop & Battle Pass &~\deficon~ I feel tempted to buy the unique mythic skin at the last level, as \\
& & &  it has custom sound and visual effects. \\ 
& & In-Game Shop& ~\deficon~ I feel tempted to buy cute items. Being able to use them make  \\
& & & gameplay more attractive. \\ \midrule
\multicolumn{2}{l}{\cellcolor{lightgray!50}\textit{Theme 6: Forced Action}} & \cellcolor{lightgray!50} & \cellcolor{lightgray!50} \\
Gamification & Pay-to-Play~\faGamepad & Battle Pass &~\deficon~ I feel pressured to pay a premium fee every season for rewards.\\
& & & ~\deficon~I feel pressured to purchase levels to unlock reward instantly\\
& & &  when I lack time for XP grinding.\\
& & New character & ~\deficon~ I feel pressured to buy the premium Battle Pass as I want to\\
& & &  unlock new character immediately.\\
 & Grinding~\faGamepad & Daily/Weekly/Seasonal Challenges &~\deficon~ I feel pressured to play the game everyday to grind the 3 (daily),\\
 & & &  11 (weekly), and 40+ (seasonal) repetitive tasks like ``earn 10 \\
 & & & eliminations/assists without dying.'' \\
 & & Battle Pass &~\deficon~ I feel pressured to grind for XPs to complete levels (10,000 XP) \\
 & & & each from gameplay and completing daily/weekly challenges. \\
 & & New Character &~\deficon~ I feel pressured to grind for unlocking the new character\\
 & & & to avoid grinding character-specific challenges in future seasons.\\
 & Playing by Appointment~\faGamepad & Double XP Weekend &~\deficon~I feel an urgency to participate in this weekend event from \\
 & & &  March 10 to 12, 2023 to grind XPs faster.\\
Sunk Cost Fallacy & Endowed Progress~\faGamepad & Battle Pass & ~\deficon~Once I have made progress, it is hard for me to abandon an\\
& & &  incomplete level and an incomplete Battle Pass.\\
& & Daily/Weekly/Seasonal Challenges & ~\deficon~ Once I have made progress, it is hard for me to abandon an \\
& & &incomplete challenge or stop before completing all daily, weekly,\\
& & &  and seasonal challenges. \\
 & Endowed Value~\faGamepad & Competitive Ranks &~\deficon~ I feel pressured to play daily to maintain my gained skills\\
 & && and character mastery.\\
 & Waste Aversion~\faGamepad & In-Game Shop &~\deficon~ I feel pressured to buy additional currency packs to use up my \\
 & & &   remaining coins from previous items and bundles purchases. \\ \midrule
\multicolumn{2}{l}{\cellcolor{lightgray!50}\textit{Theme 7: Facilitation}} & \cellcolor{lightgray!50} & \cellcolor{lightgray!50} \\
Infinite Treadmill~\faGamepad &\multicolumn{1}{c}{-}& Battle Pass &~\deficon~ I feel pressured to complete a new Battle Pass each season. \\
& & Daily/Weekly/Seasonal Challenges &~\deficon~ I feel pressured to complete new challenges daily, weekly, and\\
& & &  seasonally.\\ \midrule
\multicolumn{2}{l}{\cellcolor{lightgray!50}\textit{Theme 8: Other}} & \cellcolor{lightgray!50} & \cellcolor{lightgray!50} \\
Complete the Collection~\faGamepad &\multicolumn{1}{c}{-}& Character Item Gallery &~\deficon~ I feel the urge to complete character item collections due to a\\ 
& & &  progress bar indicating the proportion of missing items. \\
\bottomrule
\multicolumn{4}{l}{\begin{tabular}[c]{@{}l@{}}\textit{Note}. $a$. Contents in quotes are descriptions directly from the game. \\
~\faGamepad~denotes deceptive designs mentioned in the game-related literature~\cite{zagal2013dark,king20233d}. ~\faConnectdevelop~denotes deceptive designs mentioned in VR-related literature~\cite{hadan2024deceived}.\\ \end{tabular}}\\
\end{tabular}%
}
\vspace{-15mm}
\end{table*}

\begin{table*}[!ht]
\centering
\vspace{-4mm}
\caption{Based on our \textit{VR Deceptive Game Design Assessment Guide} in~\autoref{tab:initial-codebook}, this table presents deceptive design patterns we identified within \Moss and \BS.}
\vspace{-4mm}
\label{tab:VR-finalcodebook}
\resizebox{0.95\textwidth}{!}{%
\begin{tabular}{@{}llll@{}}
\toprule
\textbf{\textit{Theme}/Subtheme} & \textbf{Code} & \textbf{VR Game Mechanics$^a$} & \textbf{Synthesized Diary Entries$^b$} \\ \midrule
\multicolumn{2}{l}{\cellcolor{lightgray!50}\textit{\textbf{Theme 2: Obstruction}}} & \cellcolor{lightgray!50} & \cellcolor{lightgray!50} \\
Creating Barriers & Price Comparison Prevention &\cellcolor{paleblue!20}Beat Saber: Song Purchase System &~\cellcolor{paleblue!20}\textcolor{paleblue}{\deficon}~I feel frustrated because comparing individual song and\\ 
& &\cellcolor{paleblue!20} & \cellcolor{paleblue!20}song pack prices in-game is impossible, as no clear prices\\
& & \cellcolor{paleblue!20}& \cellcolor{paleblue!20}provided in game.\\ \midrule
\multicolumn{2}{l}{\cellcolor{lightgray!50}\textit{\textbf{Theme 3: Social Engineering}}} & \cellcolor{lightgray!50} & \cellcolor{lightgray!50} \\
Social Proof  & Parasocial Pressure &\cellcolor{paleblue!20}Beat Saber: Song Gallery & ~\textcolor{paleblue}{\deficon}~\cellcolor{paleblue!20}I feel attracted to songs and music packs from my favorite\\
& & \cellcolor{paleblue!20}& \cellcolor{paleblue!20}singers, games, or anime.\\
Urgency  & Limited Time Messages~\faGamepad &\cellcolor{paleblue!20}Beat Saber: Song Purchase System & ~\cellcolor{paleblue!20}\textcolor{paleblue}{\deficon}~I feel tempted to buy songs and music packs during the\\
& & \cellcolor{paleblue!20}&  \cellcolor{paleblue!20}time-limited sale.\\
Personalization & Competition~\faGamepad &\cellcolor{paleblue!20}Beat Saber: Leaderboard &~\cellcolor{paleblue!20}\textcolor{paleblue}{\deficon}~ While my skills may not compete with global or local\\ 
& & \cellcolor{paleblue!20}& \cellcolor{paleblue!20}players and my friends, I feel pressured to perform better to\\
& & \cellcolor{paleblue!20}& \cellcolor{paleblue!20}beat my old scores to avoid losing gained skills.\\ \midrule
\multicolumn{2}{l}{\cellcolor{lightgray!50}\textit{\textbf{Theme 4: Interface Interference}}} & \cellcolor{lightgray!50} & \cellcolor{lightgray!50} \\
Manipulating Choice Architecture & False Hierarchy~\faGamepad &\cellcolor{paleblue!20}Beat Saber: Main Menu &~\cellcolor{paleblue!20}\textcolor{paleblue}{\deficon}~I feel tempted to tap on the music pack beside game\\
& & \cellcolor{paleblue!20}& \cellcolor{paleblue!20}mode options on the main menu because it seems special \\
& & \cellcolor{paleblue!20}& \cellcolor{paleblue!20}compared to those within each game mode.\\
& Visual Prominence~\faGamepad &\cellcolor{paleblue!20}Beat Saber: Song Purchase System &~\cellcolor{paleblue!20}\textcolor{paleblue}{\deficon}~ I feel distracted by the notification that purchasing the \\
& & \cellcolor{paleblue!20}& \cellcolor{paleblue!20}entire music pack is cheaper, and the visually prominent \\
& & \cellcolor{paleblue!20}& \cellcolor{paleblue!20}``buy music pack'' option on top of the ``buy level (song)'' \\
& & \cellcolor{paleblue!20}& \cellcolor{paleblue!20}option that I intend to select. \\
 & Bundling~\faGamepad &\cellcolor{paleblue!20}Beat Saber: Song Purchase System &~\cellcolor{paleblue!20}\textcolor{paleblue}{\deficon}~ I feel buying whole music packs because an in-game\\
 & & \cellcolor{paleblue!20}& \cellcolor{paleblue!20}notification claims they are cheaper, despite no clear prices \\
 & & \cellcolor{paleblue!20}& \cellcolor{paleblue!20}listed for individual songs or packs. \\
Emotional/Sensory Manipulation & Adorable design~\faGamepad~\faConnectdevelop &\cellcolor{paleblue!20}Beat Saber: Song Gallery & ~\cellcolor{paleblue!20}\textcolor{paleblue}{\deficon}~I feel attracted to play the game repeatedly because of its \\
& & \cellcolor{paleblue!20}& \cellcolor{paleblue!20}beautiful and visually striking block and lightsaber designs.\\
& & \cellcolor{paleblue!20}& ~\cellcolor{paleblue!20}\textcolor{paleblue}{\deficon}~ I feel attracted to the BTS music pack$^c$ because each song\\
& & \cellcolor{paleblue!20}& \cellcolor{paleblue!20}has the singers' avatars dancing with me in the game. \\
& &  \cellcolor{paleblue!20}& ~\cellcolor{paleblue!20}\textcolor{paleblue}{\deficon}~ I feel attracted to songs and music packs with cute covers.\\
& & \cellcolor{paleblue!20}& ~\cellcolor{paleblue!20}\textcolor{paleblue}{\deficon}~ I feel attracted to songs with strong beats as they help\\
& & \cellcolor{paleblue!20}& \cellcolor{paleblue!20}me catch the rhythm and perform better in the game.\\
& &\cellcolor{palered!20}Moss: Book II: Character Design &~\cellcolor{palered!20}\textcolor{palered}{\deficon}~ I feel attracted to play the game by the cute character\\
& & \cellcolor{palered!20}& \cellcolor{palered!20}designs, and colorful and vibrant map environments.\\
& & \cellcolor{palered!20}&~\cellcolor{palered!20}\textcolor{palered}{\deficon}~ I feel attracted to interact with the characters because\\
& & \cellcolor{palered!20}& \cellcolor{palered!20}she actively ``talk'' to me using sign language.\\
& NPC/Narrative Manipulation~\faGamepad~\faConnectdevelop &\cellcolor{palered!20}Moss: Book II: Game Narrative &~\cellcolor{palered!20}\textcolor{palered}{\deficon}~ I lose track of time spent in game due to deep connection\\
& &\cellcolor{palered!20} & \cellcolor{palered!20}to the story and characters from the unexpected twists in\\
& & \cellcolor{palered!20}& \cellcolor{palered!20}the storyline.\\
& &\cellcolor{palered!20}Moss: Book II: 3D Interaction and Immersion & ~\cellcolor{palered!20}\textcolor{palered}{\deficon}~I lose track of time spent in game as I feel being part\\
& & \cellcolor{palered!20}& \cellcolor{palered!20}of the story because of the interactive 3D feedback that \\
& & \cellcolor{palered!20}& \cellcolor{palered!20}allow me to flip pages of the story book, move objects in \\
& & \cellcolor{palered!20}& \cellcolor{palered!20}the environment, and ``pet,'' ``hug,'' and ``high-five'' with the \\
& & \cellcolor{palered!20}& \cellcolor{palered!20}characters.\\
Trick Questions &\multicolumn{1}{c}{-}&\cellcolor{paleblue!20}Beat Saber: Song Purchase System &\cellcolor{paleblue!20}~\textcolor{paleblue}{\deficon}~ I feel confused by the option for purchasing individual \\
& & \cellcolor{paleblue!20}& \cellcolor{paleblue!20}songs (`` buy level'') as it might refer to purchasing a difficulty \\
& & \cellcolor{paleblue!20}& \cellcolor{paleblue!20}level rather than a song. Whereas the ``buy music pack''\\
& & \cellcolor{paleblue!20}& \cellcolor{paleblue!20}option is more straightforward. \\ 
Hidden Information &\multicolumn{1}{c}{-}&\cellcolor{paleblue!20}Beat Saber: 3D Interaction and Immersion &~\cellcolor{paleblue!20}\textcolor{paleblue}{\deficon}~ I lose track of time due to deep engagement of gameplay \\
& & \cellcolor{paleblue!20}& \cellcolor{paleblue!20}and the absence of an in-game clock.\\
& & \cellcolor{palered!20}Moss: Book II: 3D Interaction and Immersion&~\cellcolor{palered!20}\textcolor{palered}{\deficon}~ I lose track of time due to deep engagement of gameplay \\
& & \cellcolor{palered!20}& \cellcolor{palered!20}and the absence of an in-game clock.\\
& &\cellcolor{paleblue!20}Beat Saber: Gameplay&~\cellcolor{paleblue!20}\textcolor{paleblue}{\deficon}~I feel frustrated by the lack of an option or progress bar \\
& & \cellcolor{paleblue!20}& \cellcolor{paleblue!20}to mark specific parts of a song where I struggle. As a result, \\
& & \cellcolor{paleblue!20}& \cellcolor{paleblue!20}when I fail a game, I must repeatedly practice the entire \\
& & \cellcolor{paleblue!20}& \cellcolor{paleblue!20}song. \\ 
& &\cellcolor{palered!20}Moss: Book II: Progress Indicator &~\cellcolor{palered!20}\textcolor{palered}{\deficon}~I feel confused about my storyline progress because the \\
& & \cellcolor{palered!20}& \cellcolor{palered!20}progress indicator is too subtle and blends into the library \\
& &\cellcolor{palered!20} & \cellcolor{palered!20}environment seamlessly. \\ \midrule
\multicolumn{2}{l}{\cellcolor{lightgray!50}\textit{\textbf{Theme 6: Forced Action}}} & \cellcolor{lightgray!50} & \cellcolor{lightgray!50} \\
Forced Communication/Disclosure & Privacy Zuckering &\cellcolor{paleblue!20}Beat Saber: Song Purchase System &~\cellcolor{paleblue!20}\textcolor{paleblue}{\deficon}~I feel vulnerable because purchasing songs or music \\
& & \cellcolor{paleblue!20}& \cellcolor{paleblue!20}packs requires entering payment information before seeing \\
& & \cellcolor{paleblue!20}& \cellcolor{paleblue!20}clear prices. \\
Gamification & Pay-to-Play~\faGamepad &\cellcolor{paleblue!20}Beat Saber: Song Purchase System &~\cellcolor{paleblue!20}\textcolor{paleblue}{\deficon}~I feel annoyed by the fact that even though I bought the \\
& & \cellcolor{paleblue!20}& \cellcolor{paleblue!20}game I have to consistently buy new songs and music packs.\\
& & \cellcolor{paleblue!20}& ~\cellcolor{paleblue!20}\textcolor{paleblue}{\deficon}~I lose track of money spent on songs and music packs \\
& & \cellcolor{paleblue!20}& \cellcolor{paleblue!20}since each costs only around \$2 to \$4 USD.\\
& Power Creep~\faGamepad &\cellcolor{paleblue!20}Beat Saber: Gameplay &~\textcolor{paleblue}{\deficon}~ \cellcolor{paleblue!20}Over time, I feel bored by playing the same songs \\
&  &\cellcolor{paleblue!20} & \cellcolor{paleblue!20}repeatedly, which drives me to buy new ones. \\
 & Can't Pause or Save~\faGamepad &\cellcolor{palered!20}Moss: Book II: Saving Options &~\cellcolor{palered!20}\textcolor{palered}{\deficon}~ I feel pressured to play continuously and stop less, as the \\
& & \cellcolor{palered!20}& \cellcolor{palered!20}game solely on automatic saves. I cannot save my progress \\
& & \cellcolor{palered!20}& \cellcolor{palered!20}manually even when my battery is dying. \\
Sunk Cost Fallacy & Endowed Value ~\faGamepad &\cellcolor{paleblue!20}Beat Saber: Gameplay &~\cellcolor{paleblue!20}\textcolor{paleblue}{\deficon}~I feel pressured to play daily to maintain my gained skills \\
& & \cellcolor{paleblue!20}& \cellcolor{paleblue!20}and song mastery. \\
\multicolumn{2}{l}{\cellcolor{lightgray!50}\textit{\textbf{Theme 8: Other}}} & \cellcolor{lightgray!50} & \cellcolor{lightgray!50} \\
Complete the Collection~\faGamepad &\multicolumn{1}{c}{-}&\cellcolor{paleblue!20}Beat Saber: Song Gallery &~\cellcolor{paleblue!20}\textcolor{paleblue}{\deficon}~ I feel the urge to own all songs in a music pack due to \\
& & \cellcolor{paleblue!20}& \cellcolor{paleblue!20}my positive experiences with some songs.\\  
& &\cellcolor{palered!20}Moss: Book II: Scrolls and Armors &~\cellcolor{palered!20}\textcolor{palered}{\deficon}~ I feel the urge to collect all scrolls and armors, even \\
& &\cellcolor{palered!20} & \cellcolor{palered!20}revisiting maps, though they don't affect the storyline.\\
\bottomrule
\multicolumn{4}{l}{\begin{tabular}[c]{@{}l@{}}\textit{Note}. $a$. Mechanics from a same game are grouped by color. We used pink for game mechanics from \Moss and blue for those from \BS. \\
$b$. Contents in quotes are descriptions directly from the game. $c$. BTS Music Pack. Beat Saber Wiki. ~\url{https://beatsaber.fandom.com/wiki/BTS_Music_Pack}. Last accessed on May 4, 2024.\\
~\faGamepad~denotes deceptive designs mentioned in the game-related literature~\cite{zagal2013dark,king20233d}. ~\faConnectdevelop~denotes deceptive designs mentioned in VR-related literature~\cite{hadan2024deceived}.\\ \end{tabular}}\\
\end{tabular}%
}
\vspace{-15mm}
\end{table*}

\end{document}